# Edge Control of Graphene Domains Grown on Hexagonal Boron Nitride

*Lingxiu Chen,[a,b,f,g] Haomin Wang,[a,f]\* Shujie Tang,[a] Li He,[a,c] Hui Shan Wang,[a,d] Xiujun Wang,[a,f,g] Hong Xie,[a,f] Tianru Wu,[a,f] Hui Xi[a,e] Tianxin Li[e] and Xiaoming Xie [a,b,f]\**

a. State Key Laboratory of Functional Materials for Informatics, Shanghai Institute of Microsystem and Information Technology, Chinese Academy of Sciences, Shanghai 200050, China
b. School of Physical Science and Technology, ShanghaiTech University, Shanghai 200031, China
c. School of Optical and Electronic Information, Huazhong University of Science and Technology, Wuhan 430074, China
d. School of Physics and Electronics, Central South University, Changsha 410083, China
e. National Laboratory for Infrared Physics, Shanghai Institute of Technical Physics, Chinese Academy of Sciences, Shanghai 200083, China
f. CAS Center for Excellence in Superconducting Electronics, Chinese Academy of Sciences, Shanghai 200050, China
g. School of Electronic, Electrical and Communication Engineering, University of Chinese Academy of Sciences, Beijing 100049, China

**Abstract:** Edge structure of graphene has a significant influence on its electronic properties. However, control over the edge structure of graphene domains on insulating substrates is still challenging. Here we demonstrate edge control of graphene domains on hexagonal boron nitride (h-BN) by modifying ratio of working-gases. Edge directions were determined with the help of both moiré pattern and atomic-resolution image obtained via atomic force microscopy measurement. It is believed that the variation on graphene edges mainly attributes to different growth rates of armchair and zigzag edges. This work demonstrated here points out a potential approach to fabricate graphene ribbons on h-BN.

*\* Electronic mail: hmwang@mail.sim.ac.cn, xmxie@mail.sim.ac.cn*

**Introduction**

Graphene is well known as a two dimensional (2D) material with fantastic physical properties, such as high carrier mobility, atomically thin thickness, ultra-high mechanical strength.[1,2] The edges of graphene always affect its electronic properties, especially in its nanostructures.[3-5] Edges, as the most reactive part of graphene domain, can easily be influenced during chemical vapor deposition (CVD) processes. CVD is a prevalent approach to synthesize graphene in large scale, and has exhibited high flexibility to suppress the nucleation density,[6] enlarge single-crystal size,[7] control the number of layer[8] and modify edge structures.[9] So far, the investigation on edge control of graphene only was carried out on metal surface.[9] The edge control of graphene on insulator has not been studied.

In this work, we demonstrate the edge control of graphene on hexagonal boron nitride (h-BN) by modifying the ratio of ethyne (acting as carbon precursor) and silane (serving as gaseous catalyst)[10] during growth process of CVD. Edge orientation of graphene domains can be deduced through moiré pattern and atomic-resolution images of atomic force microscopy (AFM). Finally, graphene ribbons with different edge-directions from the step of top layer are obtained on h-BN.

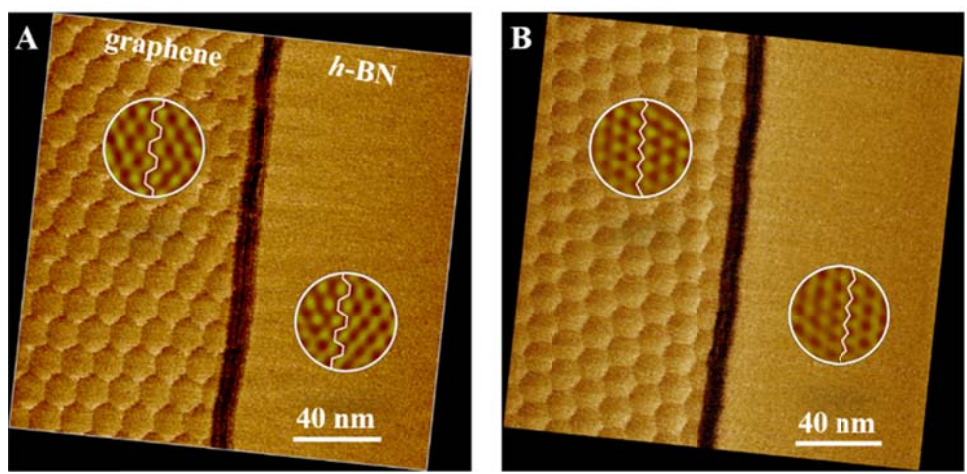

**Figure 1| Oriented edges of graphene grown on h-BN.** Graphene edge along (A) armchair and (B) zigzag direction. The circular insets show atomic-resolution AFM images, confirming the crystallographic direction of edges. The atomic-resolution images in the insets are Fourier filtered for clarity. The alignment of moiré pattern with respect to the h-BN lattice confirms that the graphene lattice is precisely aligned with underlying h-BN lattice.

**Results and discussion**
**Edge structure and moiré pattern**
AFM was adapted to investigate graphene domains grown on h-BN. Fig. 1 shows two typical AFM images of oriented edges: along armchair (AC) and zigzag (ZZ) directions. As shown in Fig. 1A, the giant super-lattice on graphene, also known as moiré patterns are observed. The appearance of the uniform regular hexagonal patterns is due to lattice mismatch between graphene and h-BN, also indicating the high quality of graphene crystallinity. Periodicity of the moiré pattern is about 13.9 nm, indicating that graphene is precisely aligned with the underlying h-BN.[11-13] The alignment of graphene domains also can be recognized by identifying the rotation of the moirépattern with respect to underlying h-BN lattice.[14] The alignment is very sensitive to the separation angle between graphene and h-BN when the misalignment angle is less than 1°. Because the misalignment of moiré patterns with respect to the h-BN is less than 3° as shown in Fig. 1A, it confirms again that the graphene is precisely aligned with the underlying h-BN. As shown in the circular insets of Fig. 1A, the atomic-resolution AFM images also indicate that the edge of graphene is along AC direction. Similarly, Fig. 1B shows that the orientation of moiré pattern is parallel to graphene edge. And with the help of atomic-resolution images in the circular insets, it can be confirmed that the edge of graphene is along ZZ direction.

**Edge control of graphene domains**

In earlier reports, hydrogen is found to play a crucial role in determination of the shape of graphene domains grown both on h-BN[15,16] and on metal[17] by acting as an etching reagent. By greatly increasing partial pressure of hydrogen during growth, edges of graphene grains become mainly along ZZ direction.[18] For the catalyst-free approaches, control over the shape, crystallinity and growth rates of graphene domains are rather poor, even after optimizing growth temperature and pressure. A precise control of the grain shape and crystallinity on h-BN requires either a plasma-enhanced approach[19] or gaseous catalyst assisted growth.[10] Our recent progress in silane-assisted growth exhibits its advance in growth of single-crystal graphene domains,[10] where Si element acts as a carbon activator while hydrogen plays as etching element. As such, silane can play a crucial role in the determination of the shape and size of graphene domains. From the point of technological view, it is more flexible to tune the ratio of ethyne ($C_2H_2$) and silane ($SiH_4$) in order to control over the shape of graphene domains and their edges.

Fig. 2A shows a graphene domain grown in 8 sccm ethyne-flow. Lattice directions (both AC and ZZ) were indicated in straight lines to tell the edge direction. It is found that the edges of the domains in Fig. 2A are mainly along AC direction. When the gas ratio of ethyne to silane was tuned to 8:1, the vertexes of hexagonal graphene domain obtained became rounded (Fig. 2B). With a further increase of silane-flow from 1 sccm to 8 sccm, the topography of graphene domains changed from hexagon with "pure" AC-edges, to dodecagon which was approximate circle, and finally to hexagon with "pure" ZZ-edges (Fig. 2E), shown in Fig. 2B-E. When the gas ratio was 6:8, the dodecagonal graphene domains are obtained with angle of interval vertexes smaller than that of their neighbours (Fig. 2F), where the orientation of moiré pattern and edges proved that the graphene domain was very close to hexagon with "pure" ZZ-edges. And with the flow of ethyne decreasing from 8 sccm to 1 sccm, the topography of graphene domains changes from hexagon with "pure" ZZ-edges (Fig. 2E) to dodecagon with high-ratio of ZZ segments (Fig. 2F), to regular dodecagon (Fig. 2G), to dodecagon with high-ratio of AC segments (Fig. 2H), and finally to hexagon with "pure" AC-edges (Fig. 2I). As shown in Fig. 2F-H, the edges of dodecagonal domains are of mixed sides with both AC and ZZ segments. More details of the growth rates and discussion are shown in Supporting Information Fig. S2-S3 and Tab. S1.

In order to investigate the topographical variation of graphene domain, a series of experiments were carried out by changing the ratio of ethyne to silane. Firstly, we maintained the flow rate of ethyne at 8 standard-state cubic centimeter per minute (sccm) and changed the flow rate of silane to 0, 1 sccm, 4 sccm, 6 sccm and 8 sccm, respectively. After that, we only decreased the ethyne flow and carried out the growth with the ratios of ethyne to silane at 6:8, 4:8, 2:8 and 1:8, respectively. Graphene domains obtained under these conditions were measured by AFM, and the topographies of these domains are shown in Fig. 2. The graphene edge structures were deduced from the orientation of moiré pattern, and verified via measurement of the lattice orientation.

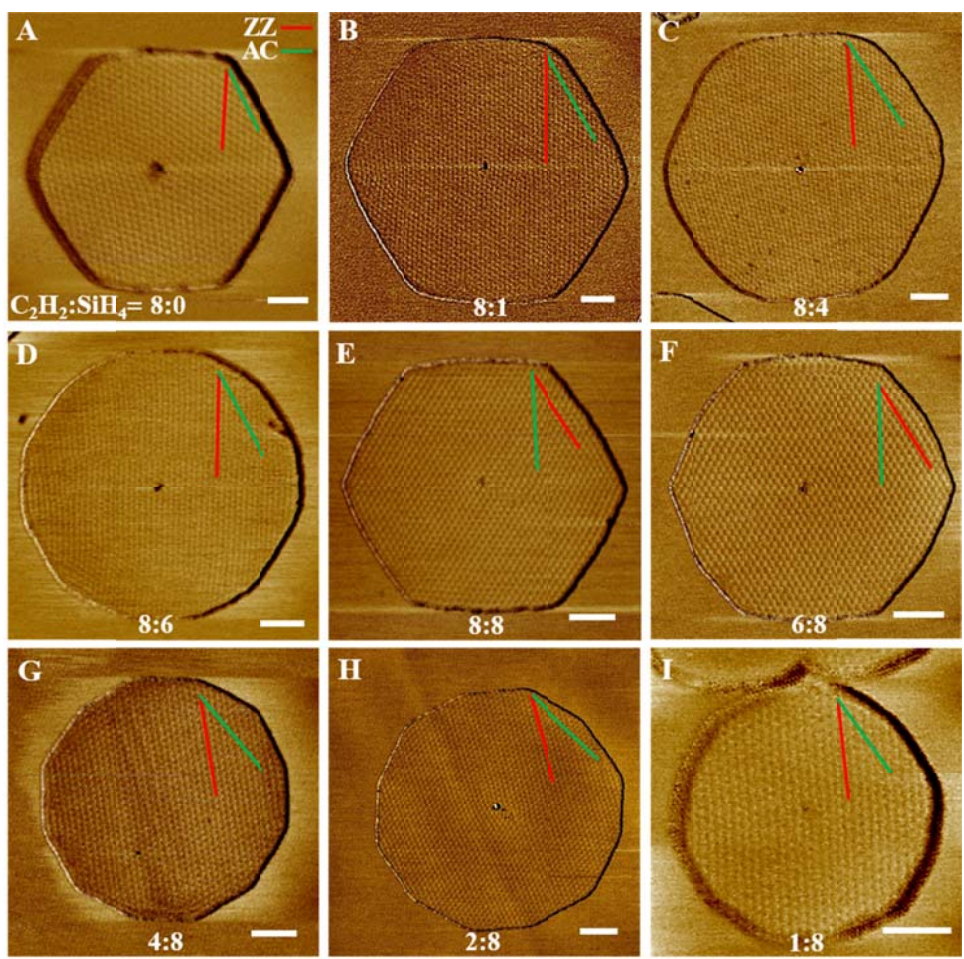

**Figure 2| Influence of gas ratio on topography of graphene domains.** Graphene domains obtained by supplying ethyne at 8 sccm and silane at (A) 0 sccm, (B) 1 sccm, (C) 4 sccm, (D) 6 sccm, and (E) 8 sccm, respectively; And graphene domains obtained by keeping silane-flow at 8 sccm and ethyne-flow at (E) 8 sccm, (F) 6 sccm, (G) 4 sccm, (H) 2 sccm, and (I) 1 sccm. The red and green lines represent lattices orientation along ZZ and AC, respectively. The scale bars are 100 nm.

In order to understand the influence of growth time on edge structures of graphene domains, we adjusted the growth time of graphene without changing any other parameter. Fig. S4 (see supporting information) shows AFM images of graphene domains obtained after a growth of 4 mins and 8 mins in a typical case. As shown in Fig. S4, the edge orientation of the domain keeps unchanged even after the growth time is doubled. Other cases show similar trends. Therefore, we believe that the edge structures of the domains remain unchanged during the growth after the domains nucleate.

Finally, we demonstrated the growth of graphene ribbons on h-BN. Besides defective sites, step-edges on h-BN surface can also serve as nucleation center.[16] The step-edges on h-BN surface can be created via mechanical exfoliation or thermal-etching. Graphene ribbons with smooth edges can be grown along the oriented step-edges of h-BN by optimizing CVD condition. As shown in Fig. 3A, graphene ribbons were grown along AC-oriented h-BN step-edges under the growth condition for graphene domains with AC-oriented edges. Fig. 3B shows a graphene ribbon with ZZ-oriented edges, which were grown along h-BN ZZ-oriented step-edges. The distortion of moiré pattern near h-BN/graphene in-plane boundary in Fig. 3A&3B is mainly due to the in-plane bonding and the lattice mismatch of graphene and h-BN.

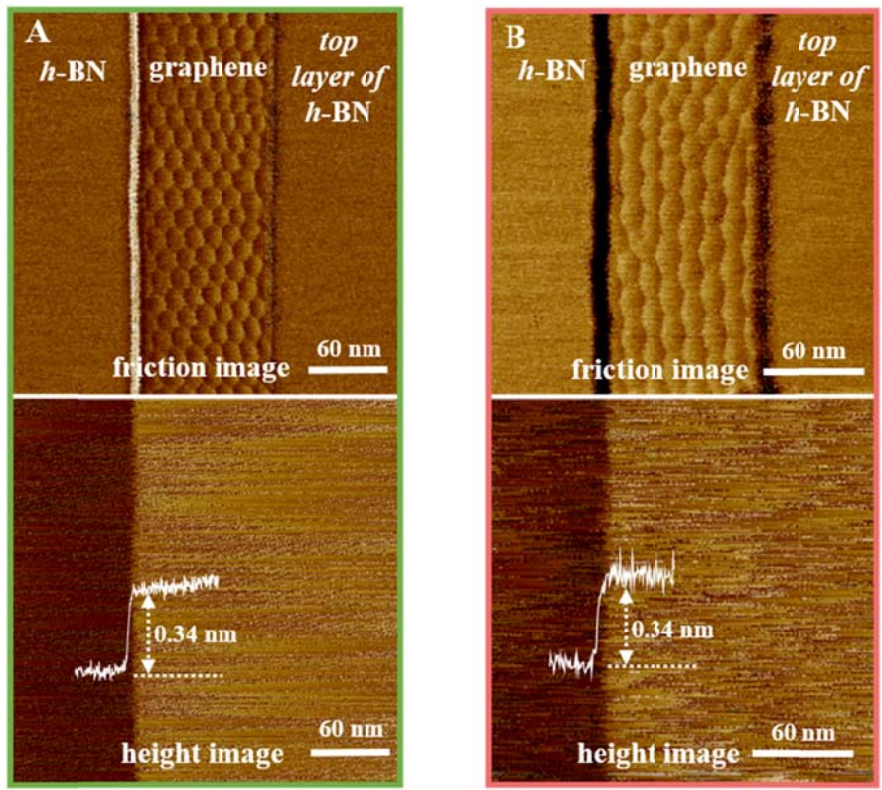

**Figure 3| Epitaxial graphene ribbons grown from oriented h-BN step-edges.** Graphene ribbons with (A) AC-oriented edges and (B) ZZ-oriented edges are grown from oriented step-edges on h-BN. The inset in height image indicates the graphene ribbon is in the thickness of monolayer.

In order to investigate the electronic properties of graphene ribbons on h-BN, a four-probe device is fabricated on 200 nm-wide AC ribbon on Si/SiO$_2$ (300 nm) substrate, and 10/50 nm Pd/Au is applied as contact metal. Fig. 4A shows the back-gate dependence of resistance at different temperature. The carrier mobility is ~5100 cm$^2$V$^{-1}$s$^{-1}$ at 300 K, indicating the high quality of epitaxial graphene ribbon. The appearance of two-side-peaks especially at 4 K is due to the existence of secondary Dirac cone (SDC) which is caused by the presence of moiré pattern.[20,21] Fig. 4B shows the magneto-resistance of the same ribbon at 4 K, showing the resistance enhancement of SDC peaks is very obvious. The Quantum Hall Effect of graphene is also seen with a filling factor ν=±2 of valleys shows in the fan diagram.

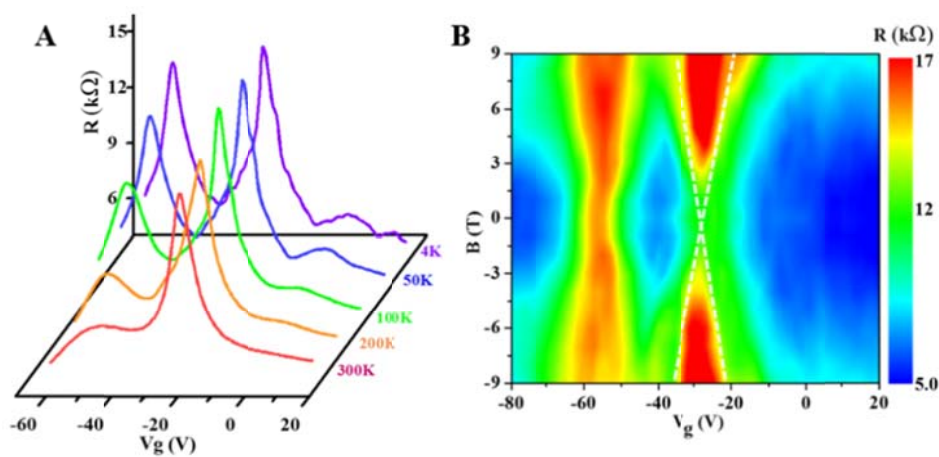

**Figure 4| Electronic transport of epitaxial graphene ribbon on h-BN.** Back-gate dependence of four-probe resistance of 200 nm-wide epitaxial graphene ribbon at different (A) temperatures and (B) magnetic field at 4 K.

## Conclusions

We demonstrate the controllable growth of graphene domains with different edge structure, and the edge of graphene domains can be controlled by modifying ethyne and silane flow in CVD process. Graphene ribbons with oriented edges were obtained from the step-edges of h-BN in optimized conditions. This work provides a promising way to fabricate graphene nano-ribbon with oriented edge if the width can be narrowed further.

## Experimental section

Before graphene growth, h-BN flakes are mechanically exfoliated onto quartz substrates, and subsequently annealed for an hour at 600 °C with oxygen flow to remove the organic residues and contaminations. The graphene growth was carried out by applying ethyne and silane (which is a mixture of argon 95% and silane 5%) flow at 1300 °C. The heating and cooling processes were performed under the protection of argon flow. More details of the experiments show in Supporting Information Fig. S1. The ordinary AFM scanning in contact mode were carried out on Bruker Dimension Icon AFM while atomic-resolution images were measured on Veeco multimode IV.

## Acknowledgements

The work performed at the Shanghai Institute of Microsystem and Information Technology, Chinese Academy of Sciences, was partially supported by the National Science and Technology Major Projects of China (Grant No. 2011ZX02707), the Chinese Academy of Sciences (Grant Nos. KGZD-EW-303, XDB04030000 and XDB04040300), a projects from the Science and Technology Commission of Shanghai Municipality (Grant No. 16ZR1442700). The work conducted at the Shanghai Institute of Technical Physics, Chinese Academy of Sciences, was partially supported by the National Science Foundation of China (Grant No. 91321311).